\input{aipcheck}

\documentclass[
    ,final            
  ]
  {aipproc}

\usepackage{epsfig,graphicx,bbm,psfrag,amssymb}
\usepackage{fancyhdr}

\layoutstyle{8x11double}

\begin{document}

\title{LFV in semileptonic $\tau$ decays and $\mu-e$ conversion in
  nuclei in SUSY-seesaw}

\classification{{11.30.Hv}, {12.60.Jv}, {14.60.St} , {{\bf Prepint
numbers:} FTUAM-08/19, IFT-UAM/CSIC-08-59}}

\author{E. Arganda}{
  address={Departamento de F\'{\i }sica Te\'{o}rica 
and Instituto de F\'{\i }sica Te\'{o}rica, IFT-UAM/CSIC \\
Universidad Aut\'{o}noma de Madrid,
Cantoblanco, E-28049 Madrid, Spain}
}

\author{M. Herrero}{
  address={Departamento de F\'{\i }sica Te\'{o}rica 
and Instituto de F\'{\i }sica Te\'{o}rica, IFT-UAM/CSIC \\
Universidad Aut\'{o}noma de Madrid,
Cantoblanco, E-28049 Madrid, Spain}
}

\author{J. Portol\'es}{
  address={IFIC, Universitat de Val\`encia - CSIC, Apt. Correus 22085, E-46071 Val\`encia, Spain}
}

\author{A. Rodr\'{\i }guez-S\'anchez}{
  address={Departamento de F\'{\i }sica Te\'{o}rica 
and Instituto de F\'{\i }sica Te\'{o}rica, IFT-UAM/CSIC \\
Universidad Aut\'{o}noma de Madrid,
Cantoblanco, E-28049 Madrid, Spain}
}

\author{A.M. Teixeira}{
  address={Laboratoire de Physique Th\'eorique, UMR 8627, 
      Paris-Sud XI, B\^atiment 210, F-91405 Orsay Cedex, France}
}


\begin{abstract}
Here we review the main results of LFV in the semileptonic tau decays $\tau \to \mu PP$ ($PP = \pi^+
\pi^-, \pi^0 \pi^0, K^+ K^-, K^0 \bar{K}^0$), $\tau \to \mu P$ ($P =
\pi, \eta, \eta^{\prime}$), and $\tau \to \mu V$ ($V = \rho, \phi$)
as well as in $\mu-e$ conversion in nuclei within SUSY-seesaw scenarios, and
compare our predictions with the present experimental
bounds\footnote{Talk given at the SUSY08 conference, Seoul, by M. Herrero.}.
\end{abstract}

\maketitle


\section{Framework for LFV}

We work within the framework of the Minimal Supersymmetric Standard Model (MSSM) enlarged by three right
handed neutrinos and their SUSY partners, where potentially observable 
LFV effects in the charged lepton sector are expected to occur. 
We further assume a seesaw mechanism
for neutrino mass generation and use the parameterisation 
$m_D =\,Y_\nu\,v_2 =\,\sqrt {m_N^{\rm diag}} R \sqrt {m_\nu^{\rm diag}}U^{\dagger}_{\rm
MNS}$, with
$R$ defined by $\theta_i$
($i=1,2,3$);  $v_{1(2)}= \,v\,\cos (\sin) \beta$, $v=174$ GeV; 
$m_{\nu}^\mathrm{diag}=\, \mathrm{diag}\,(m_{\nu_1},m_{\nu_2},m_{\nu_3})$ denotes the
three light neutrino masses, and  
$m_N^\mathrm{diag}\,=\, \mathrm{diag}\,(m_{N_1},m_{N_2},m_{N_3})$ the three heavy
ones. $U_{\rm MNS}$ is given by
the three (light) neutrino mixing angles $\theta_{12},\theta_{23}$ and $\theta_{13}$, 
and three phases, $\delta, \phi_1$ and $\phi_2$. With this 
parameterisation it is easy to accommodate
the neutrino data, while leaving room for extra neutrino mixings (from the right
handed sector). It further allows for large
Yukawa couplings $Y_\nu \sim \mathcal{O}(1)$ by
choosing large entries in $m^{\rm diag}_N$ and/or $\theta_i$. 

Here we focus in the particular LFV proccesses: 1) semileptonic $\tau \to
\mu PP$ ($PP = \pi^+ \pi^-, \pi^0 \pi^0, K^+ K^-, K^0 \bar{K}^0$), $\tau \to
\mu P$ ($P =
\pi, \eta, \eta^{\prime}$), $\tau \to
\mu V$ ($V = \rho, \phi$) decays and 2) $\mu-e$ conversion in heavy nuclei. The predictions 
in the following are for two different constrained MSSM-seesaw scenarios, 
with universal and non-universal Higgs soft masses. 
The respective parameters (in addition to the
previous neutrino sector parameters) are: 
1) CMSSM-seesaw: $M_0$, $M_{1/2}$, $A_0$ $\tan \beta$, and sign($\mu$), and 
2) NUHM-seesaw: $M_0$, $M_{1/2}$, $A_0$ $\tan \beta$, sign($\mu$),
$M_{H_1}=M_0(1+\delta_1)^{1/2}$ and
$M_{H_2}=M_0(1+\delta_2)^{1/2}$. The predictions presented here for
the $\mu-e$ conversion rates include 
the full set of SUSY one-loop contributing diagrams, mediated by $\gamma$, Z,
 and Higgs bosons, as well as boxes, and do not use the Leading
 Logarithmic (LLog) nor the mass insertion approximations.
In the case of semileptonic tau decays we have not included the boxes
which are clearly subdonimant.
The hadronisation of quark bilinears is performed within the
chiral framework, using Chiral Perturbation Theory and Resonance Chiral
Theory. This is a very short summary of the works
in~\cite{Arganda:2007jw} and~\cite{Arganda:2008jj} to which we
refer the reader for more details.

\section{Results and discussion}

Here we present the predictions for BR($\tau \to \mu PP$) ($PP = \pi^+
\pi^-, \pi^0 \pi^0, K^+ K^-, K^0 \bar{K}^0$), BR($\tau \to \mu P$) ($P =
\pi, \eta, \eta^{\prime}$), BR($\tau \to \mu V$) ($V = \rho, \phi$) and
CR($\mu-e$, Nuclei) within the previously described framework and
compare them with the following experimental bounds: BR$(\tau \to \mu \pi^+
\pi^-) < 4.8 \times 10^{-7}$, BR$(\tau \to \mu K^+
K^-) < 8 \times 10^{-7}$, BR$(\tau \to \mu \pi) < 5.8 \times 10^{-8}$,
BR$(\tau \to \mu \eta) < 5.1 \times 10^{-8}$, BR$(\tau \to \mu
\eta^{\prime}) < 5.3 \times 10^{-8}$, BR$(\tau \to \mu \rho) < 2
\times 10^{-7}$, BR$(\tau \to \mu \phi) < 1.3 \times 10^{-7}$,
CR$(\mu-e, {\rm Au}) < 7 \times 10^{-13}$ and CR$(\mu-e, {\rm Ti}) <
4.3 \times 10^{-12}$.

\begin{figure}
\begin{tabular}{c}
\psfig{file=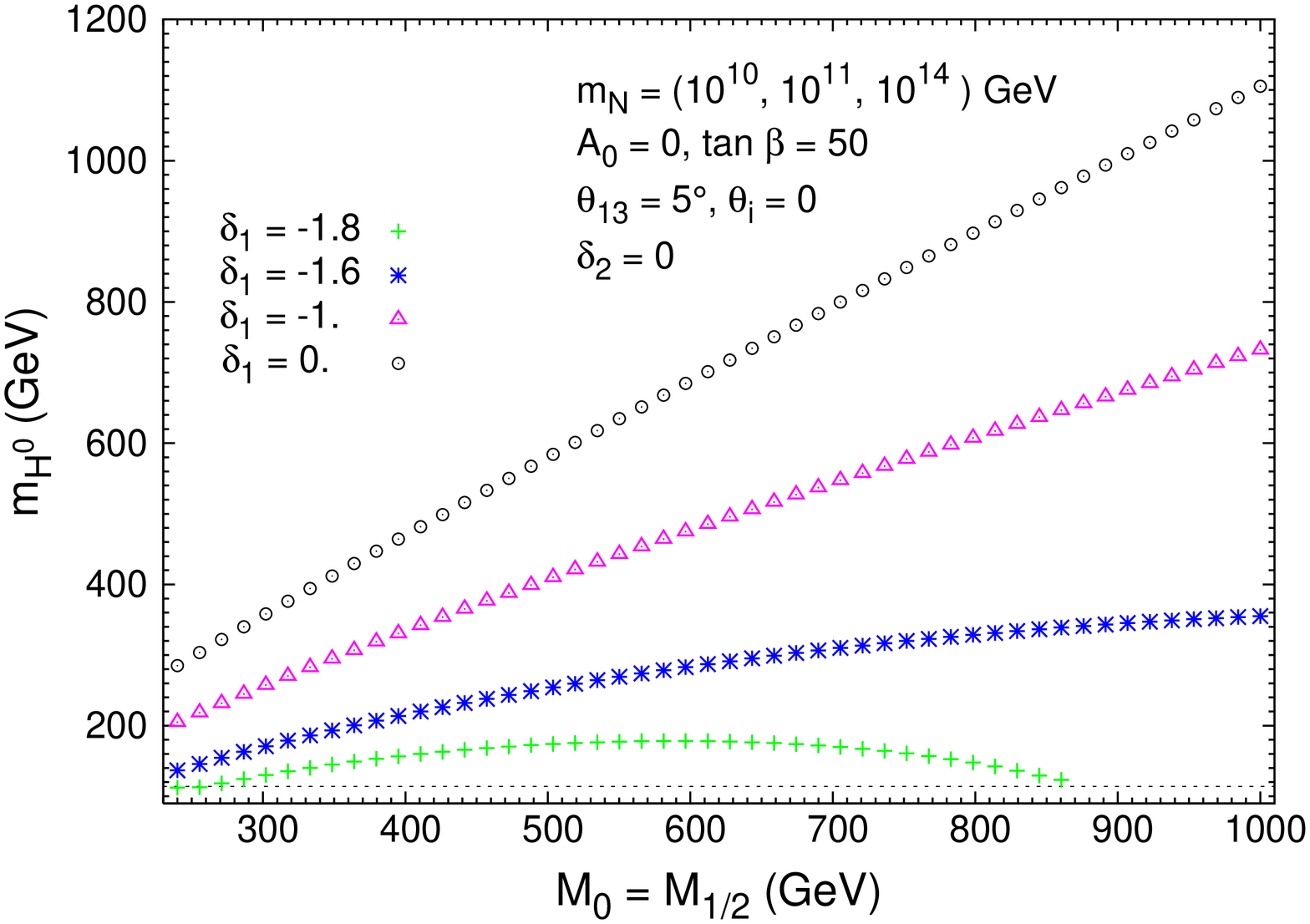,width=50mm,angle=270,clip=}\\
\psfig{file=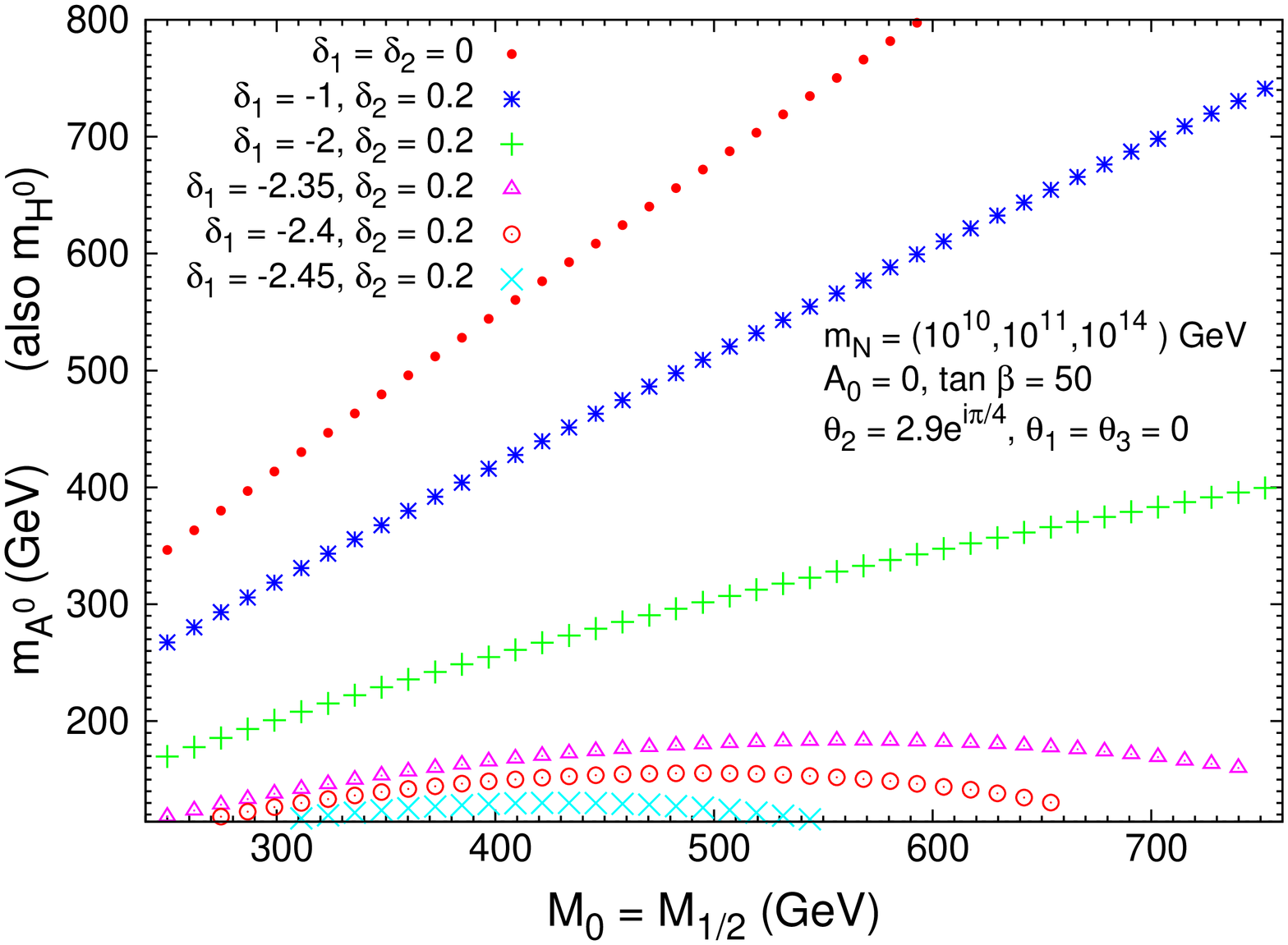,width=50mm,angle=270,clip=}
\caption{Predictions of the Higgs boson masses as a function of
 $M_{\rm SUSY}$ in CMSSM ($\delta_1 = \delta_2 = 0$) and NUHM.}
\label{fig:1}       
\end{tabular}
\end{figure}

As a general result in LFV processes that can be mediated by Higgs
bosons we have found that the $H^0$ and $A^0$ contributions are relevant at
large $\tan\beta$ if the Higgs masses
are light enough. It is in this aspect where the main difference between the two
considered scenarios lies. Within the CMSSM, light Higgs $H^0$ and $A^0$
bosons are only possible for low $M_{\rm SUSY}$ (here we take $M_{\rm SUSY} =
M_0 = M_{1/2}$ to reduce the number of input parameters). In
contrast, within the NUHM, light Higgs bosons can be obtained even at
large $M_{\rm SUSY}$. In Fig.~\ref{fig:1} it is shown that some
specific choices of $\delta_1$ and $\delta_2$ lead to values of
$m_{H^0}$ and $m_{A^0}$ as low as 110-120 GeV, even for  heavy $M_{\rm SUSY}$
values above 600 GeV. Therefore, the sensitivity to the Higgs sector
is higher in the NUHM.

\begin{figure}
\begin{tabular}{c}
\psfig{file=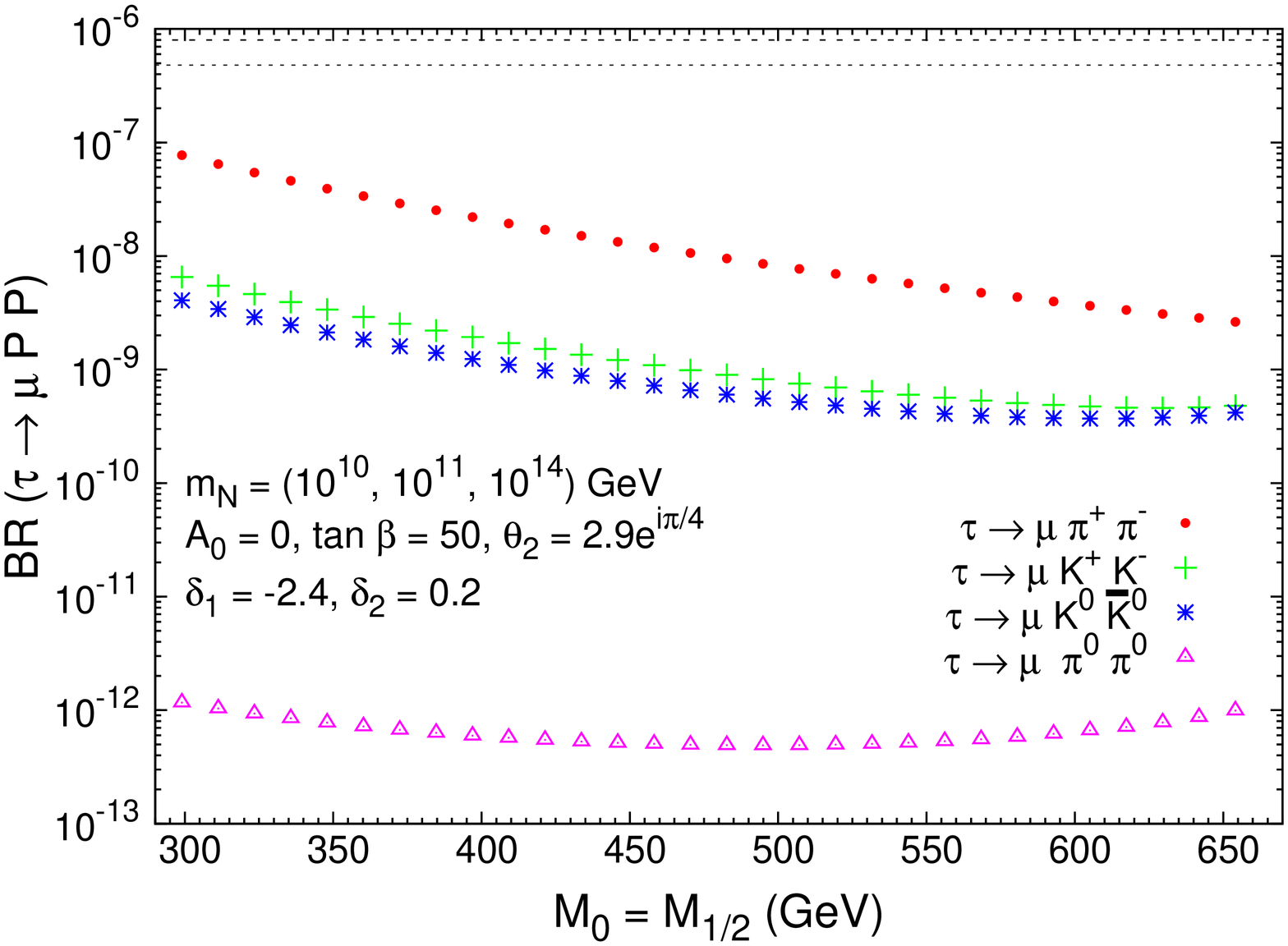,width=50mm,angle=270,clip=}\\
\psfig{file=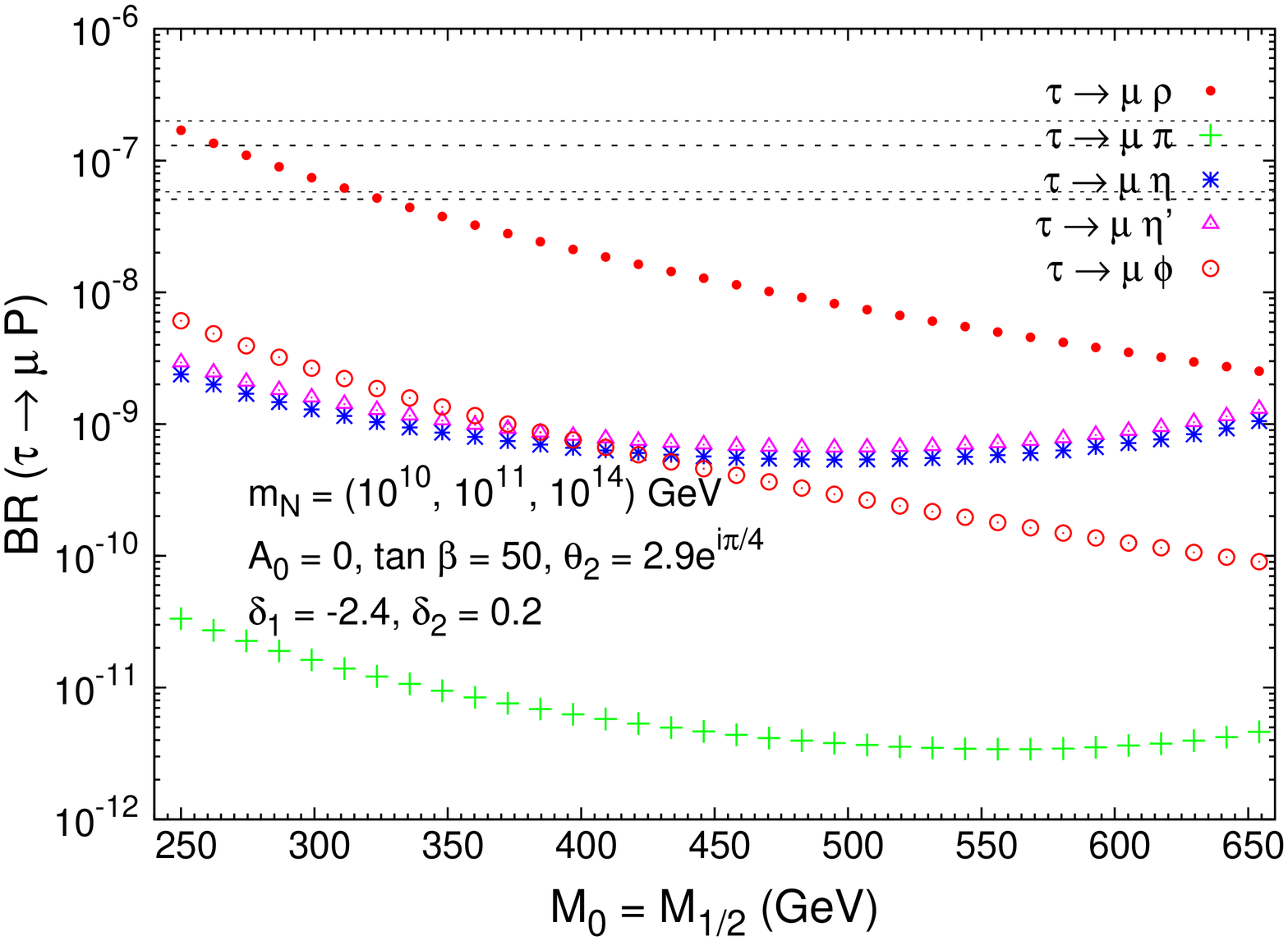,width=50mm,angle=270,clip=}
\caption{Present sensitivity to LFV in semileptonic $\tau$ decays
within the NUHM scenario. The horizontal lines denote
experimental bounds.}
\label{fig:2}       
\end{tabular}
\end{figure}

We start by presenting the results for the semileptonic tau
decays. The mentioned sensitivity to the Higgs sector within the NUHM
scenario can be seen
in Fig.~\ref{fig:2}. Concretely, the BRs of the channels $\tau \to \mu K^+ K^-$,
$\tau \to \mu K^0 \bar K^0$, $\tau \to \mu \pi^0 \pi^0$, $\tau
\to \mu \pi$, $\tau \to \mu \eta$ and $\tau \to \mu \eta^{\prime}$
present a growing behaviour with $M_{\rm SUSY}$, in the large $M_{\rm
SUSY}$ region, due to the contribution of light Higgs bosons,
which is non-decoupling. The decays involving Kaons and $\eta$
mesons are particularly sensitive to the Higgs contributions because
of their strange quark content, which has a stronger coupling to the Higgs bosons.
On the other hand, the largest predicted
rates are for $\tau \to \mu \pi^+ \pi^-$ and $\tau \to \mu \rho$,
dominated by the photon contribution, which are indeed at the present experimental
reach in the low $M_{\rm SUSY}$ region.

\begin{figure}
\psfig{file=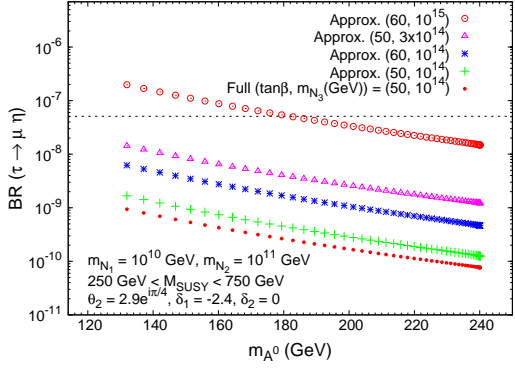,width=50mm,angle=270,clip=}
\caption{Sensitivity to Higgs sector in $\tau \to \mu \eta$ decays.}
\label{fig:3}       
\end{figure}

The maximum sensitivity to the Higgs sector is found for $\tau \to \mu
\eta$ and $\tau \to \mu \eta^{\prime}$ channels, largely dominated by
the $A^0$ boson exchange. Fig.~\ref{fig:3} shows that BR($\tau \to \mu
\eta$) reaches the experimental bound for large heaviest neutrino mass,
large $\tan\beta$, large $\theta_i$ angles and low $m_{A^0}$. For the choice of input
parameters in this figure, it occurs at 
$m_{N_3} = 10^{15}$ GeV, $\tan\beta = 60$, $\theta_2 = 2.9
e^{i\pi/4}$ and $m_{A^0} = 180$ GeV.
A set of useful formulae for all these channels, within the mass
insertion approximation which are
valid at large $\tan\beta$, are presented in~\cite{Arganda:2008jj}.
We have shown that the predictions with these formulae agree with the
full results within a factor of about 2. In the case of $\tau \to \mu
\eta$ this comparison is shown in
Fig.~\ref{fig:3}. Similar conclusions are found for $\tau \to \mu
\eta^{\prime}$. The next relevant channel in sensitivity to the Higgs
sector is $\tau \to \mu K^+ K^-$, but
it is still below the present experimental bound. To our knowledge,
there are not experimental bounds yet  available for $\tau \to \mu K^0
\bar K^0$ and $\tau \to \mu \pi^0 \pi^0$.

\begin{figure}
\begin{tabular}{cc}
\psfig{file=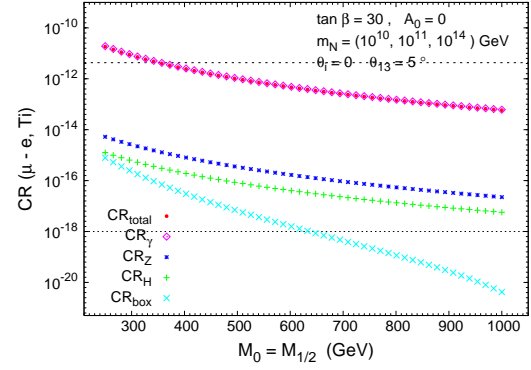,width=50mm,angle=270,clip=}\\
\psfig{file=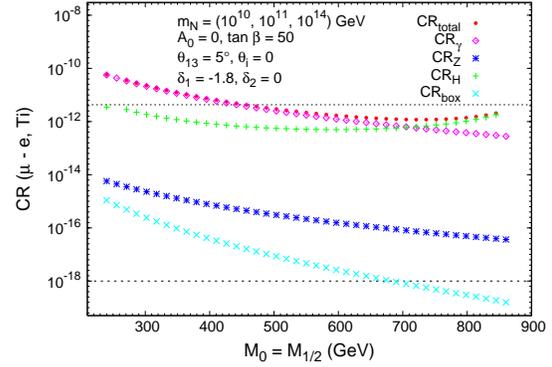,width=50mm,angle=270,clip=}
\caption{Predictions of CR($\mu-e$, Ti) as a function of
 $M_{\rm SUSY}$ in the CMSSM (above) and NUHM (below) scenarios.}
\label{fig:4}       
\end{tabular}
\end{figure}

Next we comment on the results for $\mu-e$ conversion in
nuclei. Fig.~\ref{fig:4} shows our predictions of the conversion rates
for Titanium as a function of $M_{\rm SUSY}$ in both CMSSM and NUHM
scenarios. As in the case of semileptonic tau decays, the sensitivity
to the Higgs contribution is only manifest in the NUHM scenario. The
predictions for CR($\mu-e$, Ti) within the CMSSM scenario are largely
dominated by the photon contribution and present a decoupling
behaviour at large $M_{\rm SUSY}$. In this case the present
experimental bound is only reached at low $M_{\rm SUSY}$. The
perspectives for the future are much more promising. If the announced
sensitivity by PRISM/PRIME of $10^{-18}$ is finally attained, the full studied
range of $M_{\rm SUSY}$ will be covered. 

Fig.~\ref{fig:4} also illustrates that within the NUHM scenario the
Higgs contribution dominates at large $M_{\rm SUSY}$ for light Higgs
bosons. The predicted rates are close to the present experimental bound
not only in the low $M_{\rm SUSY}$ region but also for heavy SUSY
spectra. As in the previous semileptonic tau decays, we have found in
addition a simple formula for the conversion rates, within the mass
insertion approximation, which is
valid at large $\tan\beta$~\cite{Arganda:2007jw} and can be used for
further analysis.

\begin{figure}
\psfig{file=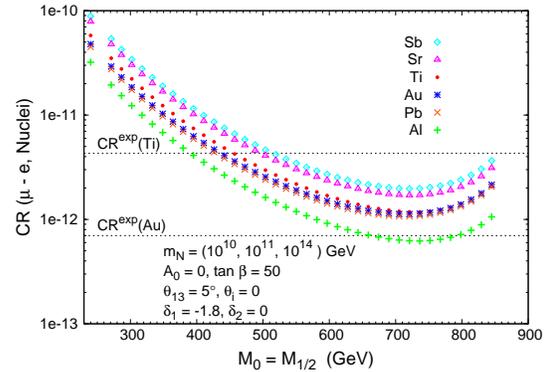,width=50mm,angle=270,clip=}
\caption{Present sensitivity to LFV in $\mu-e$ conversion for several
nuclei within NUHM.}
\label{fig:5}       
\end{figure}

The predictions of the $\mu-e$ conversion rates for several nuclei are
collected in Fig.~\ref{fig:5}. We can see again the growing behaviour
with $M_{\rm SUSY}$ in the large $M_{\rm SUSY}$ region due to the
non-decoupling of the Higgs contributions. At present, the most
competitive nucleus for LFV searches is Au where, for the choice of
input parameters in this figure, all the predicted rates are above the
experimental bound. We have also shown in~\cite{Arganda:2007jw} that $\mu-e$ conversion in
nuclei is extremely sensitive to $\theta_{13}$, similarly to $\mu \to
e \gamma$ and $\mu \to 3e$ and, therefore, a future
measurement of this mixing angle can help in the searches of LFV in
the $\mu-e$ sector.

In conclusion, we have shown that semileptonic tau decays nicely complement
the searches for LFV in the $\tau-\mu$ sector, in addition to
$\tau \to \mu \gamma$. The future prospects for $\mu-e$ conversion in
Ti are the most promising for LFV searches. Both processes,
semileptonic tau decays and $\mu-e$ conversion in nuclei are indeed more
sensitive to the Higgs sector than $\tau \to 3 \mu$. 

\begin{theacknowledgments}
M. Herrero acknowledges the organisers for her
invitation to give this talk and for the fruitful conference.   
\end{theacknowledgments}

\end{document}